\documentclass[twocolumn,prc,aps,floatfix]{revtex4}
\usepackage[dvips]{graphicx}
\begin{document}
\title{Limits on cosmological variation of quark masses and strong interaction}
\author{V.F. Dmitriev$^{1,2}$ and V.V. Flambaum$^{1,3}$}
\affiliation{$^1$
 School of Physics, The University of New South Wales, Sydney NSW
2052, Australia
}
\affiliation{$^2$
Budker Institute of Nuclear Physics, 630090, Novosibirsk-90,
Russia
}
\affiliation{$^3$  Institute for Nuclear Theory,
 University Of Washington, Seattle, WA 98195-1550, USA
}
\date{\today}
\begin{abstract}
We discuss limits on variation of
$(m_q/\Lambda_{QCD})$.
The results are obtained by studying  $n-\alpha$-interaction
during Big Bang, Oklo natural nuclear reactor
data and limits on varition of the proton
$g$-factor from quasar absorpion spectra.
\end{abstract}
\maketitle

PACS number: 98.80.Cq

\section{Introduction}
Recent astronomical data suggest a possible variation of the fine
structure constant $\alpha=e^2/\hbar c$ at the $10^{-5}$ level over a
time-scale of 10 billion years, see \cite{alpha} (a discussion of
other limits can be found in Ref. \cite{Brax} and references therein).
The data motivated immediately more general discussions of possible
variations of other constants. Unlike for the electroweak forces for
the strong interaction there is generally no direct relation between
the coupling constants and observable quantities. In a recent paper
\cite{FS} we presented general discussions of  possible influence
of the strong scale variation on primordial Big Bang Nucleosynthesis
(BBN) yields, the Oklo natural nuclear reactor, quasar absorption
spectra and atomic clocks.

 Here we continue this work, concentrating
on the properties of $^5$He (BBN), Sm
(Oklo natural nuclear reactor),  $^{12}$C (stars)
 and proton magnetic g-factor (quasar
absorption spectra). Since one can measure only variation of the
dimentionless quantities, we can extract from our results variation
of the dimensionless ratio $m_q/\Lambda_{QCD}$ where $m_q$ is the quark
mass (with the dependence on the normalization point removed)
 and $\Lambda_{QCD}$ is the strong interaction scale determined
by a position  of the pole in perturbative QCD running coupling constant.
It is convinient to assume that $\Lambda_{QCD}$ is  constant, and quark
mass $m_q$ is  variable.

   A position of an energy level in a strong potential depends both
on the parameters of the strong interaction and nucleon mass $M$.
For example, in a simplest case of a particle in a deep square potential
well the single-particle energy  levels are
\begin{equation} \label{E}
E_n \approx -V + \frac{\hbar^2 \pi^2 (n+l/2)^2}{2 M R^2},
\end{equation}
where $V$ and $R$ are the depth and radius of the potential.
The strong interaction is sensitive to the light quark mass because
the $\pi$-meson mass $m_\pi \sim \sqrt{m_q \Lambda_{QCD}}$,
where $m_q=m_u+m_d$.
The nucleon mass is also sensitive to quark mass because $M$ has large
 ``strange fraction'' \cite{sigmaterm}
\begin{equation} \label{Mstrange}
{\partial M \over \partial m_s}=<N|\bar s s |N>\approx 1.5
\end{equation}
Putting  the strange quark mass $m_s=130 \,\,MeV$ one finds that
 about 1/5 of the nucleon
 mass comes from the ``strange
 term''.

\section{n - $\alpha$ interaction}

The production of nuclei with $A>5$ during BBN is strongly suppressed
because of the absence of stable nuclei with $A=5$.
$^5$He is unstable nucleus which is seen as a resonance in n-$\alpha$
elastic scattering at neutron laboratory energy around 1.1 MeV \cite{S68}.
The resonance corresponds to the ground state of unstable nucleus $^5$He with
mass excess  $\Delta =11386.234$ KeV \cite{nndc}. The ground state lies at
0.89 MeV above neutron threshold. This energy is rather small in nuclear scale,
therefore, rather small variation of the strong interaction scale may influence
considerably the position of the resonance making, for example, $^5$He stable.
Stable $^5$He at the time of BBN would change strongly the primordial abundances
of light elements. Basing on the fact that the standard BBN theory describes
rather well the observed light element abundances we can put bounds on strong
scale variation at the time of BBN.

A phenomenological potential describing $n-\alpha$ scattering phases at low
energy has the Woods-Saxon shape \cite{S68}.
\begin{equation} \label{pot}
U(r) = -V(e^x+1) +(\hbar/m_\pi c)^2V_s({\bf L}\cdot\mbox{\boldmath $\sigma$})
r^{-1}(d/dr)(e^{x_s}+1). \end{equation}
In Eq.(\ref{pot}),
\begin{equation} \label{x}
  x = (r-R)/a, \;\;\; x_s = (r-R_s)/a_s.
\end{equation}
Here, $R$ and $R_s$ are the central and the spin-orbit radiuses of the
potential,
$$
R= r_0 (M_\alpha/M_p)^{1/3}, \;\;\; R_s= r_s (M_\alpha/M_p)^{1/3},
$$
where $(M_\alpha/M_p)=3.973$ is the ratio of the $\alpha$ and the proton
masses. The other parameters of the potential are
$$
\begin{array}{ll}

V = 41.88\; {\rm MeV}, &\;\; V_s = (3.0 +0.1 E_n)\; {\rm MeV} \nonumber \\
r_0=(1.50 -0.01 E_n)\; {\rm fm}, &\;\;r_s =1.0\; {\rm fm} \nonumber \\
a = a_s = 0.25\; {\rm fm}&

\end{array}
$$
The main contributions to this potential come from exchange of scalar and
vector mesons. These contributions are not related directly to the pion and we
can expect them to be insensitive to variation of the pion properties.
Pion exchange interaction gives zero contribution into the
potential via direct term. The exchange term, however, gives nonzero contribution. If we
isolate this contribution, we can study, at least approximately,  the dependence
of $^5$He ground state properties on pion mass and pion-nucleon coupling
constant.
\subsection{One pion exchange (OPE) contribution}
We start from the standard OPE interaction
\begin{equation}  \label{ope}
V_\pi({\bf r}_1 -{\bf r}_2)= -\frac{f^2}{\mu_0^2}(\mbox{\boldmath
$\tau$}_1\cdot\mbox{\boldmath $\tau$}_2)(\mbox{\boldmath
$\sigma$}_1\cdot\mbox{\boldmath $\nabla$}_1) (\mbox{\boldmath
$\sigma$}_2\cdot\mbox{\boldmath $\nabla$}_2)e^{-m_\pi r_{12}}/r_{12}.
\end{equation}
Here we distinguish between pion mass in the exponent and pion mass in the
denominator. The last one is used for normalization of pion-nucleon coupling
constant and should not be varied.

The exchange contribution into $n-\alpha$ potential can be written in the form
\begin{equation}   \label{exch}
 W({\bf r}_1,{\bf r}_2) =
-\frac{3}{4\pi}\frac{f^2}{\mu_0^2}m_\pi^2R_0(r_1)\frac{1}{r_{12}}e^{-m_\pi
r_{12}}R_0(r_2).
\end{equation}
In Eq.(\ref{exch}), $R_0(r)$ is 1s radial wave function of nucleon inside
$\alpha$-particle. For our purpose it is sufficient to choose $R_0(r)$ as a
simple gaussian
\begin{equation}
R_0(r)= \frac{1}{b^{3/2}}(\frac{2}{\pi})^{1/4}\exp(-r^2/4b^2),
\end{equation}
where the parameter $b$ is related to mean-square size of the
$\alpha$-particle
$$
\overline{r^2} = 3b^2.
$$
We omitted the contact term in Eq.(\ref{exch}). It does not depend on the pion
mass and can be included into phenomenological part of the potential
Eq.(\ref{pot}) together with other mesons contribution. With the potential
Eq.(\ref{exch}) the Schr\"odinger equation for $n-\alpha$ system becomes
nonlocal. For $p_{3/2}$ radial wave function $\chi_1(r)$ we obtain the equation
\begin{eqnarray} \label{radeq} -\frac{\hbar^2}{2m}\frac{d^2}{dr^2}\chi_1(r)
+(U(r) + \frac{\hbar^2}{2m}\frac{2}{r^2})\chi_1(r)-
6\frac{f^2}{\mu_0^2}\frac{m_\pi^3}{b^3}\frac{2}{\pi}& & \nonumber \\
\left[rk_1(m_\pi r)e^{-r^2/4b^2}\int_0^r r'i_1(m_\pi r')e^{-r'^2/4b^2}\chi_1(r')
dr' +\right.&&\nonumber \\ \left. ri_1(m_\pi r)e^{-r^2/4b^2}\int_r^\infty
r'k_1(m_\pi r')e^{-r'^2/4b^2}\chi_1(r')dr'\right]&& \nonumber \\
= E \chi_1(r).&&
\end{eqnarray}
In Eq.(\ref{radeq}), $i_1(x)$, and $k_1(x)$ are the spherical Bessel functions
of imaginary argument.

Using Eq.(\ref{radeq}) we can now adjust the phenomenological parameter $V$
putting the resonance in $p_{3/2}$ wave in its position $E=0.89$ MeV. The
procedure gives $V= 32.74$ MeV that should be compared with 41.88 MeV for full
phenomenological potential Eq.(\ref{pot}). The difference between these two
values can be treated as an effective depth of equivalent local potential
corresponding to nonlocal exchange contribution of pion, Eq.(\ref{exch}).
The Eq.(\ref{exch}) was obtained neglecting pion-nucleon formfactor. This is
reasonable approximation since the cutoff parameter $\Lambda$ in the formfactor
is large compared to nucleon momentum in $\alpha$-particle nucleus.

\subsection{BBN bounds on pion, nucleon and quark  mass variation}
Using Eq.(\ref{radeq}) we can study the dependence of $p_{3/2}$ resonance
position on pion mass.  The potential Eq.(\ref{exch}) is proportional to pion
mass squared. Therefore, an increase in pion mass would lead to lowering of the
resonance position. Simple numerical exercise leads to conclusion that the
increase in pion mass on 21.9\% would produce $^5$He nucleus with zero neutron
binding energy. Any further increase of pion mass would lead to stable $^5$He
producing serious modifications in the process of nucleosynthesis.
Using the relation for
the $\pi$-meson mass $m_\pi \sim \sqrt{m_q \Lambda_{QCD}}$ we obtain the limit
\begin{equation}\label{heqlimit}
\frac{\delta (m_q/\Lambda_{QCD})}{(m_q/\Lambda_{QCD})} < 0.4
\end{equation}
Note that this OPE path to obtain the limits does not look very reliable -
see discussions in recent papers \cite{FS,savage}. Therefore, below we
 present the limits obtained in a different way.

 Variation of the nucleon mass $M$ leads to change of the kinetic energy
and change of the resonance position. We found that 5\% increase of the nucleon
 mass produces  $^5$He bound state with zero neutron binding energy.
Using the relation (\ref{Mstrange}) we find the limit for the variation
of the strange quark mass
\begin{equation}\label{heslimit}
\frac{\delta (m_s/\Lambda_{QCD})}{(m_s/\Lambda_{QCD})} < 0.25
\end{equation}
These are limits on the variations of the light quark mass and
strange quark mass from BBN to present time.
\section{Limits from Oklo natural nuclear reactor}
 Similar limits follow from data on natural nuclear
reactor in Oklo active about 2 bn years ago.
 The most sensitive phenomenon (used previously for limits on
the variations of the electromagnetic $\alpha$) is disappearance of certain
isotopes (especially $^{149}$Sm) possessing a neutron resonance close to zero
 \cite{Oklo}. Today the lowest
 resonance energy  $E_0=0.0973 \pm 0.0002 \, eV$
is larger compared to its width, so the
neutron capture  cross section $\sigma \sim 1/E_0^2$.
The data constrain the ratio of this cross section to the non-resonance
one which was used to find neutron flux. It
therefore implies ( under assumption that the same resonance
was the lowest one at the time of Oklo reactor)
 that these data constrain the variation of the following
ratio $\delta( E_0/E_1) $ where $E_1\sim 1 \, MeV$ is a typical
single-particle  energy scale which is determined mainly by $\Lambda_{QCD}$.

  A very small value of $E_0$ appears as a result of the nearly exact
cancellation of two large terms $E_0=E_c - S_n$ where $S_n$ is the
neutron separation energy in $^{150}$Sm and $E_c$ is the energy of
 the many-body excited  compound state relative to the ground state
 of $^{150}$Sm
(recall that we consider the reaction $n$ +$^{149}$Sm) . The difference
is very small because we deliberately selected the lowest compound resonance
above neutron threshold. The neutron separation energy  is
$S_n=V-\epsilon_F$
where $V$ is the depth of the potential and $\epsilon_F=p_f^2/2M \sim
(\hbar^2 A^{2/3})/(M R^2)$
 is the Fermi energy. The energy of compound state $E_c$ is approximately equal
to the difference of several single-particle energies given by Eq. (\ref{E}).
Therefore, both $\epsilon_F$ and $E_c$ scale as
$\hbar^2/ (M R^2)$. We can present the resonance energy in the
 forllowing form:
\begin{equation}\label{E0}
E_0=E_c - S_n=E_c +\epsilon_F -V= K\frac{\hbar^2}{M R^2} - V
\end{equation}
where $K$ is a numerical constant. We can find this constant
from the present time condition $E_0 \approx 0$:
\begin{equation}\label{K}
K (\frac{\hbar^2}{M R^2})_{present}= V_{present}
\end{equation}
Now we may consider change of $E_0$ under variation of $M,\,R,\,V$.
\begin{equation}\label{delta}
\delta E_0 =- K\frac{\hbar^2}{M R^2}(\frac{\delta M}{M}+\frac{2 \delta R}{R})
 -\delta V =
 -V \frac{\delta M}{M} - ( V  \frac{2 \delta R}{R} + \delta V)
\end{equation}
Comparing with the $^5He$ calculation in the previous
section we can say that two last terms in the brackets correspond
 to the variation of the energy level due to the variation of
the pion mass $m_\pi$. This gives us the following expression
\begin{equation}\label{delta1}
\delta E_0 = -V \frac{\delta M}{M} + \frac{dE}{d m_\pi} \delta m_\pi=
\frac{dE}{d M} \delta M + \frac{dE}{d m_\pi} \delta m_\pi
\end{equation}
We can find numerical values of the  derivatives using
 $V$= 50 MeV, $\delta M/M = 0.2 \delta m_s/m_s $ (see Eq. (\ref{Mstrange})),
 $\delta m_\pi/m_\pi = 0.5 \delta m_q/m_q $.
\begin{equation}\label{delta2}
\delta E_0 = -0.05 \delta M - 0.03 \delta m_\pi = -10 \,
 MeV \frac{\delta m_s}{m_s} - 2 \,
 MeV \frac{\delta m_q}{m_q}
\end{equation}
Note that in  the $^5He$ calculation
 the derivative $\frac{dE}{d M}$ had  smaller absolute value
 (-0.02 instead of -0.05 here).
The difference may be related to the fact that the $p$-wave resonance
wave function  in  $^5He$ was localised mainly outside the nucleus
(where the coefficient $V$ in Eq. (\ref{delta1}) is equal to zero).
We extrapolated the derivative $\frac{dE}{d m_\pi}$ from the  $^5He$ calculation
(accuracy in this term is very low anyway).

   The limit on the shift of the Sm resonance is  \cite{Oklo}
$|\delta E_0| < 0.02$ eV. A comparison with Eq. (\ref{delta2}) gives
very stringent limits:
\begin{equation}\label{plimit}
|\frac{\delta (M/\Lambda_{QCD})}{(M/\Lambda_{QCD})}
+0.6\frac{\delta (m_\pi/\Lambda_{QCD})}{(m_\pi/\Lambda_{QCD})}|
 < 4 \times 10^{-10}
\end{equation}
\begin{equation}\label{slimit}
|\frac{\delta (m_s/\Lambda_{QCD})}{(m_s/\Lambda_{QCD})}
+0.2\frac{\delta (m_q/\Lambda_{QCD})}{(m_q/\Lambda_{QCD})}|
 < 2 \times 10^{-9}
\end{equation}
 Note that the authors of the last work in \cite{Oklo}
found also the non-zero solution $ \delta E_0  = -0.097 \pm 0.008$ eV.
This solution corresponds to the same resonance moved below
thermal neutron energy. In this case
\begin{equation}\label{p1limit}
\frac{\delta (M/\Lambda_{QCD})}{(M/\Lambda_{QCD})}
+0.6\frac{\delta (m_\pi/\Lambda_{QCD})}{(m_\pi/\Lambda_{QCD})}
 =(2 \pm 0.2)\times 10^{-9}
\end{equation}
\begin{equation}\label{s1limit}
\frac{\delta (m_s/\Lambda_{QCD})}{(m_s/\Lambda_{QCD})}
+0.2\frac{\delta (m_q/\Lambda_{QCD})}{(m_q/\Lambda_{QCD})}=
 (1 \pm 0.1) \times 10^{-8}
\end{equation}
In principle, the total number of the solutions can be very large since
$^{150}$Sm nucleus has millions of resonances and each of them
can provide two new solutions (thermal neutron energy on the right tale
 or left tale of the resonance). However, these extra solutions are probably
 excluded by the measurements of the neutron capture cross-sections
for other nuclei since no significant changes have been observed there
also, see \cite{Oklo}.

\section{Limits from $^{12}C$ production in stars and
quasar absorption spectra}

In the previous sections we obtained limits on variation of
$m_s/\Lambda_{QCD}$ during
the interval between the Big Bang and present time and on
shorter time scale from Oklo natural nuclear reactor
which was active 1.8 billion years ago. It is also possible
to obtain limits on the intermediate time scale.
One possibility is related to  postion of the resonance
in  $^{12}C$ during production of this element in stars.
This famous resonance at E=380 KeV  is needed
to produce enough carbon and create life. According
to Ref. \cite{c12} the position of this resonance
can not shift by more than 60 KeV (
one can also find in Ref. \cite{c12} the limits
on the strong interactions and other relevant references).
 We can use Eq. (\ref{delta2})
to provide some rough estimates only:
\begin{equation}\label{cplimit}
|\frac{\delta (M/\Lambda_{QCD})}{(M/\Lambda_{QCD})}
+0.6\frac{\delta (m_\pi/\Lambda_{QCD})}{(m_\pi/\Lambda_{QCD})}|
 < 1.2 \times 10^{-3}
\end{equation}
\begin{equation}\label{cslimit}
|\frac{\delta (m_s/\Lambda_{QCD})}{(m_s/\Lambda_{QCD})}
+0.2\frac{\delta (m_q/\Lambda_{QCD})}{(m_q/\Lambda_{QCD})}| < 6 \times 10^{-3}
\end{equation}
Unfortunetely, these limits are relatively weak. Much stronger limits can
 be obtained from the measurements of  quasar absorption spectra.
Comparison of atomic H 21 cm (hyperfine) transition with molecular
rotational transitions gave the following limits on $Y\equiv\alpha^2 g_p$
\cite{Murphy1}\\ $\delta Y/Y=(-.20 \pm 0.44) 10^{-5}$ for redshift
z=0.2467
and  $\delta Y/Y=(-.16 \pm 0.54) 10^{-5}$ for
z=0.6847. The second limit corresponds to roughly t=6 billion years ago.
These results were used in \cite{Murphy1} to obtain the limits
on variation of $\alpha$.
 In Ref. \cite{FS} it was suggested to use $\delta Y/Y$
to estimate variation of  $m_q/\Lambda_{QCD}$.
Here we continue this work.

    According to  calculation
in Ref. \cite{Thomas} dependence of the proton magnetic moment
on $\pi$-meson mass $m_\pi$ can be approximated by the following equation
\begin{equation}\label{thomas}
\mu_p(m_\pi)=\frac{\mu_p(0)}{1+ a m_\pi + b m_\pi ^2}
\end{equation}
where $a$= 1.37/$GeV ,  $b$= 0.452/$GeV$^2$. The proton $g$-factor
is $g_p=\mu_p/(e \hbar/2 M c)$. Using  Eq. (\ref{Mstrange}) we obtain
the following estimate:
\begin{equation}\label{g}
\frac{\delta g_p}{g_p} =\frac{\delta M}{M}
 -0.18 \frac{\delta m_\pi}{m_\pi}=0.2\frac{\delta m_s}{m_s}
 -0.09 \frac{\delta m_q}{m_q}
\end{equation}
Comparing with the limits on $\delta Y/Y$  we obtain
for the red shift z=0.2467
 (neglecting variation of $\alpha$):
\begin{equation}\label{gM1}
\frac{\delta (M/\Lambda_{QCD})}{(M/\Lambda_{QCD})}
-0.18\frac{\delta (m_\pi/\Lambda_{QCD})}{(m_\pi/\Lambda_{QCD})}
 = (-0.2 \pm 0.44) \times 10^{-5}
\end{equation}
\begin{equation}\label{g1s}
2\frac{\delta (m_s/\Lambda_{QCD})}{(m_s/\Lambda_{QCD})}
-\frac{\delta (m_q/\Lambda_{QCD})}{(m_q/\Lambda_{QCD})}
 = (-2.0 \pm 4.4) \times 10^{-5}
\end{equation}
For the red shift  z=0.6847 :
\begin{equation}\label{gM2}
\frac{\delta (M/\Lambda_{QCD})}{(M/\Lambda_{QCD})}
-0.18\frac{\delta (m_\pi/\Lambda_{QCD})}{(m_\pi/\Lambda_{QCD})}
 = (-0.16 \pm 0.54) \times 10^{-5}
\end{equation}
\begin{equation}\label{g2s}
2\frac{\delta (m_s/\Lambda_{QCD})}{(m_s/\Lambda_{QCD})}
-\frac{\delta (m_q/\Lambda_{QCD})}{(m_q/\Lambda_{QCD})}
 = (-1.6 \pm 5.4) \times 10^{-5}
\end{equation}
There is also a  limit on $X\equiv\alpha^2 g_p m_e/M_p$
\cite{Cowie} $\delta X/X=(0.7\pm 1.1) 10^{-5}$ for z=1.8.
This limit was interpreted as a limit on variation of $\alpha$
or $m_e/M_p$.

     Using eq. (\ref{thomas}) we obtain
\begin{equation}\label{gm}
\frac{\delta (g_p m_e/M)}{(g_p m_e/M)} =\frac{\delta m_e}{m_e}
 -0.18 \frac{\delta m_\pi}{m_\pi}=\frac{\delta m_e}{m_e}
 -0.09 \frac{\delta m_q}{m_q}
\end{equation}
This gives us a limit for the red shift  z=1.8
\begin{equation}\label{gM3}
\frac{\delta (m_e/\Lambda_{QCD})}{(m_e/\Lambda_{QCD})}
-0.09\frac{\delta (m_q/\Lambda_{QCD})}{(m_q/\Lambda_{QCD})}
 = (0.7 \pm 1.1) \times 10^{-5}
\end{equation}
We should stress that in this paper we present experimental errors
only. The theoretical errors are   hard to estimate, therefore
some of these limits, possibly, should be treated as  order of magnitude
 estimates.

 One of the authors (VF) is thankful to the Institute for Nuclear
Theory, University of Washington for hospitality and support.
This work is also supported by the Australian Research
Council.  We are grateful to Michael Kuchiev
for his advice about Mathematica and
 to Arkady Vainshtein for useful discussions.

\end{document}